\documentclass{emulateapj}
\lefthead{KIM \& MORRIS}
\righthead{DYNAMICAL FRICTION NEAR THE GALACTIC CENTER}
\journalinfo{Accepted for publication in ApJ; Scheduled for Nov 1, 2003 issue.}
\submitted{}

\newcommand\msun{{\rm \,M_\odot}}

\newcommand\HeI{He~{\small I} }

\begin{document}
\title{DYNAMICAL FRICTION ON STAR CLUSTERS NEAR THE GALACTIC CENTER}
\author{Sungsoo S. Kim\altaffilmark{1}}
\affil{Kyung Hee University, Dept. of Astronomy \& Space Science,
Yongin-shi, Kyungki-do 449-701, Korea; sungsoo.kim@khu.ac.kr}
\and
\author{Mark Morris}
\affil{Department of Physics \& Astronomy, University of
California, Los Angeles, CA 90095-1562; morris@astro.ucla.edu}

\altaffiltext{1}{Also at Institute of Natural Sciences, Kyung Hee University}

\begin{abstract}
Numerical simulations of the dynamical friction suffered by a star
cluster near the Galactic center have been performed with a parallelized
tree code. Gerhard (2001) has suggested that dynamical friction, 
which causes a cluster
to lose orbital energy and spiral in towards the galactic center,
may explain the presence of a cluster of very young stars in the
central parsec, where star formation might be prohibitively
difficult owing to strong tidal forces.  The clusters modeled in our
simulations have an initial total mass of $10^5$--$10^6 \msun$ and
initial galactocentric radii of 2.5--30~pc.  We have identified
a few simulations in which dynamical friction indeed brings a cluster
to the central parsec, although this is only possible if the cluster 
is either very massive ($\sim 10^6 \msun$), or is formed
near the central parsec ($\la 5$~pc).  In both cases, the cluster
should have an initially very dense core ($> 10^6 \msun$pc$^-3$).
The initial core collapse and segregation of massive stars into the 
cluster core, which typically happens on a much shorter time scale 
than that characterizing the dynamical inspiral of the cluster toward 
the Galactic center, can provide the requisite high density.  Furthermore, 
because it is the cluster core which is most likely to survive the cluster
disintegration during its journey inwards, this can help account for the 
observed distribution of presumably massive \HeI stars in the central 
parsec.

\end{abstract}

\section{INTRODUCTION}
\label{sec:introduction}

The luminosity of the central parsec of our Galaxy is dominated by a 
cluster of very young stars, including more than a dozen very 
luminous \HeI emission-line stars (Krabbe et al. 1995; Paumard et 
al. 2001) and many O and B stars (Eckart, Ott, \& Genzel 1999).  
Krabbe et al. (1995) find that the properties of the massive,
early-type stars in the central parsec can be accounted for by a burst of 
star formation between 3 and 7~Myr ago.  The \HeI stars appear to be 
very massive ($> 40 \msun$) stars at the luminous blue variable 
phase or Wolf-Rayet stage, consistent with stellar ages of $\sim 5$~Myr 
(Paumard et al. 2001).  
A total of 16 \HeI stars are now known within the central parsec, and 
there are very few \HeI stars outside this region, other than 
in two very young star clusters at least 30 pc from the Galactic center 
(described below).  Only one emission-line star has been reported between
radii of 1 and 5~parsecs (Cotera et al. 1999).

Such extremely young ages suggest {\it in situ} formation for the cluster.
However, the very strong tidal forces there raise the question of whether
the gas density can reach the limiting Roche value.  While the maximum gas
density in the central parsec, a few times $10^6 \, {\rm cm^{-3}}$ (Jackson
et al. 1993), currently appears to be substantially lower than the minimum
density required for a cloud to remain bound, $\sim6~\times~10^7 \,
{\rm cm^{-3}}$ (Morris 1993), little can now be said about the conditions
there when the central cluster formed.  One possibility is that the
gravitational collapse leading to the formation of the present cluster
of young stars in the central parsec was triggered by infall of a
particularly dense gas cloud, which experienced compression by shocks
involving cloud-cloud collisions, self-intersecting gas streams, or
violent explosions near or at the central black hole (Morris 1993;
Sanders 1998; Morris, Ghez \& Becklin 1999).
The possibility that the \HeI stars in the central
parsec are something more exotic than massive, windy stars, such as
Thorne-Zytkow type objects, was also considered by Morris (1993),
who invoked the large stellar number density there to argue that
mergers between compact stellar remnants and normal stars might
have given rise to such a population.  However, the subsequent
finding that \HeI stars are found in other massive stellar clusters
in the Galactic center region (e.g., the Quintuplet Cluster; Figer,
McLean, \& Morris 1999) eliminated the rationale for that hypothesis.

Alternatively, the star cluster could have formed outside the central
parsec, where tidal forces are relatively weaker and star formation is
consequently less problematic, and later migrated into the Galactic
center (GC) (Gerhard 2001).  This idea is motivated by the presence of two 
other massive, young star clusters near the GC, the ``Arches cluster'' 
(Nagata et al 1995; Cotera et al. 1996; Figer et al. 1999) and the
``Quintuplet cluster'' (Okuda et al. 1990; Nagata et al. 1990;
Glass, Moneti, \& Moorwood 1990; Figer et al. 1999), 
both lying within $\sim 35$~pc, in projection, of the GC.  These clusters 
are only 2-5~Myr old and contain very luminous emission-line stars similar 
to those in the central parsec (Figer, McLean, \& Morris 1995; Figer et al.
1999).  Gerhard (2001) proposes that dynamical friction can bring a massive 
young star cluster, initially embedded in its parent molecular cloud, into 
the central parsec during the lifetime of its most massive stars, depending 
on the initial location and mass of the cluster.  The drag force represented 
by dynamical friction, acting in the direction opposite to the cluster motion,
is owed to the induced ``wake'' of background stars.  If the star cluster is 
massive enough, the resulting deceleration can in principle be large enough
to cause the cluster to spiral into the GC.

When the original Chandrasekhar formula for deceleration by dynamical 
friction (Chandrasekhar 1943) is applied to the Arches and Quintuplet
clusters, both of which are estimated to have had an initial cluster mass
of several $10^4 \msun$ (Figer et al. 1999; Kim et al. 2000), one
arrives at a timescale of $\sim 100$~Myr for them to spiral into the GC.
Even with initial cluster mass of $10^5$~Myr, the friction timescale
$\tau_{fric}$ would not be shorter than 30~Myr, which is much longer
than the lifetimes of the \HeI stars within the central parsec.  
Gerhard (2001), however, finds much smaller time scales for the cluster
migration by assuming much larger cluster masses, or by considering 
that the parent molecular cloud might stay bound to the cluster so that the
effective cluster mass is large enough to yield a suitably short dynamical
friction time scale.  He finds that a massive cluster and its associated
cloud, with combined mass of $10^6 m_6 \msun$ and galactocentric radius
of $\le 30 m_6^{1/2}$~pc will reach the central parsec within the lifetime
of the \HeI stars observed in the central parsec.  

Numerical simulations of dynamical friction for globular clusters or 
satellite galaxies in the galactic halo have shown that the 
Chandrasekhar formula provides an accurate description for the orbital 
motion of a body experiencing dynamical friction (e.g., White 1976;
Lin \& Tremaine 1983).  However, the formula describes the orbiting cluster 
as a single, rigid particle, 
while an actual star cluster has a rather smooth, extended density 
profile and gradually loses its stars beyond the tidal radius. Near the
GC, the tidal radius is small (compared to the cluster mass) enough to be
an important determinant of the cluster fate (Kim et al. 1999, 2000).
Using a simple assumption for
the density profile of the cluster (an isothermal sphere with a tidal
cutoff), Gerhard modified the Chandrasekhar formula to take this effect 
into account.  However, the behavior of stars near and outside
the tidal radius of the cluster is rather difficult to describe, and
the ``lingering time'' of those stars is found to often be much
longer than the dynamical timescale of the cluster (Fukushige \& Heggie
2000; Takahashi \& Portegies Zwart 1998).

Furthermore, the tidal radius of the cluster continuously shrinks as
the cluster approaches the GC, and the cluster will eventually 
disrupt.  The distribution of cluster members in galactocentric radius
after the disruption will be determined by the galactocentric radius 
at which the cluster disrupts, by the deviation of the cluster's 
orbital motion from circularity, and by the shape of the gravitational 
potential of the Galaxy.  The final distribution of the remnant stars 
is an important issue because most clusters will begin to disrupt before 
reaching the central parsec, and even for those which do survive as a 
separate entity and reach the central parsec, the kinetic energies of
the remnant stars after disruption may be large enough that they are
not bound inside the central parsec.
Therefore, the final fate of a cluster experiencing dynamical friction 
can only be learned using numerical simulations, which can take these
myriad and complex effects into account.

We have carried out numerical simulations of the orbital and structural
evolution of star clusters situated in a realistic Galactic potential in
order to further explore the hypothesis that star clusters can be brought
into the center by dynamical friction, to accurately determine the
timescales characterizing the process, and to predict the final
distribution of remnant stars after disruption.
We judge it likely that
the parent molecular cloud will be separated from the star cluster
relatively shortly after cluster formation\footnote{
Some mechanisms for the loss of angular momentum apply only to the gas 
component, such as cloud collisions and other viscous interactions with
the interstellar medium, as well as magnetic viscosity
(Morris \& Serabyn 1996).  Also, the radiation pressure from a massive
cluster can exert sufficient force to separate cluster and cloud on
a relatively short time scale, if the cluster is not located exactly
at the center of the cloud.
}, so we do not consider the parent cloud in our simulations.
We confirm that
it is possible for the `central parsec cluster' to have resulted from a
star cluster formed outside the central parsec and brought inwards by
dynamical friction.  However, the constraints imposed by this mechanism
appear to be rather severe: the cluster must either begin its
journey quite close to the GC, or be much more massive than any
currently observed cluster, or both.

Our models and the method of simulation are described in 
\S~\ref{sec:models}, and the simulation results are presented and
discussed in \S~\ref{sec:results}.  Our findings are then summarized
and discussed in \S~\ref{sec:discussion}.

\begin{deluxetable}{ccccc}
\tablecolumns{5}
\tablewidth{0pt}
\tablecaption{Galaxy Models
\label{table:galaxy}}
\tablehead{
\colhead{} &
\colhead{$R_{trunc}$} &
\colhead{$R_{out}$} &
\colhead{$M_{galaxy}$\tablenotemark{a}} &
\colhead{} \\
\colhead{Model} &
\colhead{(pc)} &
\colhead{(pc)} &
\colhead{($M_\odot$)} &
\colhead{$N_{galaxy}$\tablenotemark{b}}
}
\startdata
1 & 80 & 120 & $3.1 \times 10^8$ & $2   \times 10^6$ \\
2 & 15 &  25 & $4.3 \times 10^6$ & $2.8 \times 10^5$ \\
3 & 20 &  30 & $6.0 \times 10^6$ & $4   \times 10^5$ \\
\enddata
\tablenotetext{a}{Total mass of the galaxy particles}
\tablenotetext{b}{Total number of galaxy particles}
\end{deluxetable}

\section{MODELS}
\label{sec:models}

\subsection{The Code}
\label{sec:code}
For the simulations presented here, we use an N-body/SPH (Smoothed 
Particle Hydrodynamics) code named {\sc Gadget}, which is freely available at
{\sf \mbox{http\,://www.mpa-garching.mpg.de}\linebreak[0]\mbox{/gadget}}
(Springel, Yoshida, \& White 2001).  {\sc Gadget} computes gravitational
forces with a hierarchical tree algorithm and represents fluids by means
of SPH.  The code uses individual and adaptive timesteps for all particles.
We adopted a parallelized version of {\sc Gadget}, which implements the
standardized MPI (Message Passing Interface) communication interface,
and we use only the gravitational part of the code.
The code is thus effectively a gravitational N-body code, with no hydrodynamical
effects included.

\subsection{The Galaxy}
\label{sec:galaxy}

We adopt a truncated, softened, spherical, power-law density profile
for the central region of the Galaxy:
\begin{equation}
	\rho_g = {4 \times 10^6 \over 1+(R/0.17\,{\rm pc})^{1.8}} \,
		 \exp(-(R/R_{trunc})^6) \, \msun {\rm pc^{-3}}.
\end{equation}
This is a density model from Genzel et al. (1996) with an added exponential
truncation.  By introducing such truncation, rather than using a 
simple cutoff at the outer boundary, we were able to significantly 
reduce numerical inaccuracies involved in converting the density profile 
to the distribution function (see below).  Depending on the initial $R$
(galactocentric radius) of the cluster, we use values 15--80~pc for
the truncation radius, $R_{trunc}$, which are set to be 2.5 to 6 times
larger than the initial $R$ in order to ensure suitable representation
of the Galaxy particles outside the initial $R$ of the cluster.  The
outer boundary of the Galaxy, $R_{out}$, is set to be 50~\% larger
than $R_{trunc}$.  Table~\ref{table:galaxy} shows three sets of
$R_{trunc}$ and $R_{out}$ used in our simulations.

\begin{deluxetable*}{cccccccccc}
\tablecolumns{10}
\tablewidth{0pt}
\tablecaption{Simulation Parameters
\label{table:sim}}
\tablehead{
\colhead{} &
\colhead{$R$} &
\colhead{$v_{init}$\tablenotemark{a}} &
\colhead{$M_{cl}$} &
\colhead{} &
\colhead{$r_c$} &
\colhead{} &
\colhead{$\rho_c$} &
\colhead{Galaxy} &
\colhead{Galaxy} \\
\colhead{Simulation} &
\colhead{(pc)} &
\colhead{($v_{circ}$\tablenotemark{b} )} &
\colhead{($\rm M_\odot$)} &
\colhead{$N_{cl}$} &
\colhead{(pc)} &
\colhead{$r_t/r_c$} &
\colhead{($\rm M_\odot pc^{-3}$)} &
\colhead{Rotation} &
\colhead{Model}
}
\startdata
1  & 30  & 1   & $10^6$ & $10^5$ & 0.86  & 6   & $1.3 \times 10^5$ & N & 1 \\
2  & 30  & 0.5 & $10^6$ & $10^5$ & 0.86  & 6   & $1.3 \times 10^5$ & N & 1 \\
3  & 30  & 0.5 & $10^6$ & $10^5$ & 0.86  & 6   & $1.3 \times 10^5$ & Y & 1 \\
4  & 30  & 0.5 & $10^6$ & $10^5$ & 0.28  & 19  & $3.8 \times 10^6$ & N & 1 \\
5  & 5   & 0.5 & $10^6$ & $10^5$ & 0.28  & 6   & $3.8 \times 10^6$ & N & 2 \\
6  & 10  & 0.5 & $10^6$ & $10^5$ & 0.43  & 6   & $1.0 \times 10^6$ & N & 3 \\
7  & 10  & 0.5 & $10^6$ & $10^5$ & 0.28  & 9.3 & $3.8 \times 10^6$ & N & 3 \\
8  & 5   & 0.5 & $10^5$ & $10^4$ & 0.13  & 6   & $3.8 \times 10^6$ & N & 2 \\
9  & 2.5 & 0.5 & $10^5$ & $10^4$ & 0.080 & 6   & $1.6 \times 10^7$ & N & 2 \\
10 & 2.5 & 0.5 & $10^5$ & $10^4$ & 0.039 & 12  & $1.3 \times 10^8$ & N & 2 \\
11 & 2.5 & 1   & $10^5$ & $10^4$ & 0.039 & 12  & $1.3 \times 10^8$ & N & 2 \\
12 & 5   & 0.5 & $10^5$ & $10^4$ & 0.039 & 20  & $1.3 \times 10^8$ & N & 2 \\
13 & 10  & 0.5 & $10^5$ & $10^4$ & 0.039 & 30  & $1.3 \times 10^8$ & N & 3 \\
\enddata
\tablenotetext{a}{Initial cluster velocity}
\tablenotetext{b}{Circular velocity at a given $R$}
\end{deluxetable*}

For the non-rotating Galaxy model, the system is assumed to be in
equilibrium, and to have an isotropic velocity distribution.  We
thus obtain the distribution function by integrating the Eddington
formula (Binney \& Tremaine 1987).  The initial positions and velocities
of the particles representing the Galaxy are then determined from the
distribution function in a Monte Carlo fashion.

Some of our simulations implement a rotating Galaxy model.
In order to produce an equilibrated rotating Galaxy model, we simply
``flip'' the sign (take the absolute value) of the azimuthal velocity
component (that about the rotation axis) for a certain fraction of particles
whose locations and original velocities were chosen as just described.
This technique, developed by Lynden-Bell (1960),
produces a rotating system without requiring modification of either
its total potential and kinetic energies or the ratio between its
radial and tangential velocities.  Lindqvist, Habing, \& Winnberg
(1992) showed that the average radial velocity $\bar v_r$ of OH/IR
stars in the central $\sim 100$~pc of the Galaxy can be fit by a
relation $\bar v_r = 1.1 \, {\rm km/s} \, (l/{\rm pc})+c$ with an
average dispersion of $82 \, {\rm km/s}$, where $l$ is the Galactic
longitude and $c$ is a constant close to 0.  We find that the observed
rotational characteristics of the OH/IR stars can be well reproduced
by flipping the following fraction of particles, $f_{flip}$, expressed
as a function of galactocentric radius projected onto the plane normal
to the rotation axis, $R_p$:
\begin{equation}
	f_{flip} = {\rm min} \, [ \, 0.4 (R_p/30 \, {\rm pc}), \, 1 \, ].
\end{equation}
We use this fraction for our rotating Galaxy models.

Depending on $R_{trunc}$, we use $2.8 \times 10^5$, $4 \times 10^5$, or
$2 \times 10^6$ particles to model the Galaxy, but the mass represented
by a single particle is always $\sim 150 \msun$.  To model the compact,
massive object at the center of the Galaxy, one particle with mass 
$2.5 \times 10^6 \msun$ is put at the center of the system.

The Galaxy in our simulations can be regarded as a collisionless system
for the time intervals considered here.
A particle-based method like {\sc Gadget} requires softening 
of gravity to better describe collisionless systems.  However, 
determining the most appropriate softening length, $\epsilon$, for a 
given problem and method is not trivial.  If $\epsilon$ is too small, 
unphysical relaxation would take place, and if too large, a given galactic 
density distribution would not be properly represented.  There seems to be no 
universal way to find the ``optimal softening'', and each problem/method
requires a series of experiments to obtain its own optimal value (Merritt
1996; Athanassoula et al. 2000; among others).  Our experiments without
a cluster component indicate
that 0.3~pc for Galaxy model~1 and 0.1~pc for Galaxy models 2 \& 3 are
the minimum $\epsilon$ values that result in an acceptably small temporal
deviation in density profile for the region of interest and the time
interval covered.  Because the majority of particles that induce dynamical
friction on a cluster are outside the tidal radius of the cluster, and because
adopted values of $\epsilon$ in our models are always smaller than the tidal
radii of the clusters, the choice of $\epsilon$ will not significantly affect
the dynamical friction time scales that result from our models.

The experiments showed that for Galaxy model~1, the density profile remained
nearly constant over time for $R>2$~pc, and for Galaxy models 2 \& 3, the
enclosed mass did not vary by more than 10~\% everywhere, for the time
interval covered in our simulations.
A slight density enhancement inside 1~pc ($\sim 20$~\% at 1~pc) was
observed during the simulation for Galaxy models 2 \& 3, which seems
to be due to the numerical inaccuracies introduced by the large mass
ratio between the Galaxy particles and the black hole particle.  However,
in that central region, the uncertainty is dominated by the estimation
of the density profile in that region, so the density enhancement
should not affect the scientific results given in the present paper.
We set the $\epsilon$ value of the black hole particle to be the same as
the Galaxy particles.

\subsection{The Cluster}
\label{sec:cluster}
 
We initially model the cluster with the Plummer density profile,
\begin{equation}
	\rho_{cl} = {M_{cl} \over 4 \pi r_c^3} \left ( 1 +
		   {r^2 \over r_c^2} \right )^{-5/2},
\end{equation}
where $M_{cl}$ is the cluster mass and $r_c$ the cluster core 
radius.  The locations and velocities of cluster particles are chosen using
this profile in the same manner as for the Galaxy particles.

Clusters in our simulations have a total mass of either $10^5 \msun$ or
$10^6 \msun$.  We set a cluster particle to represent $10 \msun$,
so the number of particles for a cluster is $10^4$ or $10^5$.

{\sc GADGET} is not designed to handle details of the internal dynamics of
star clusters such as close encounters between stars/binaries, and thus it
cannot accurately describe the dynamical evolution of star clusters.  The
dynamical evolution may be important for our lower mass ($10^5 \msun$)
clusters, and we will discuss this more \S\ref{sec:discussion})
We find that $\epsilon$ of 0.025~pc is an appropriate softening length 
for the cluster particles.  This value
keeps our clusters approximately stationary for the time intervals
investigated.  This softening value is applied only between cluster particles.
For interactions between particles representing different components,
{\sc Gadget} uses the larger of the two softening lengths (for our models,
$\epsilon$ for the galaxy particles is always larger than that for the
cluster particles).

Although we use a fairly large number of particles for the Galaxy,
Poisson noise could be problematic at large $R$ or in regions of low $\rho_g$ 
(the average distance between particles of Galaxy model~1 at $R=30$~pc 
is $\sim 0.9$~pc).  For example, since we model the cluster as a system 
of particles, rather than as a point source, Poisson noise in the Galaxy 
potential could conceivably heat the cluster and affect its internal 
structure.  However, we find that during its first revolution period, our 
cluster with $M_{cl} = 10^6 \msun$, in a circular orbit located at $R=30$~pc
(the largest $R$ among our simulations), maintained a roughly steady
density profile and mean velocity dispersion (within a few percent).
We thus believe that the effect of Poisson noise from our Galaxy
model on the cluster is negligible.

\section{RESULTS}
\label{sec:results}

\subsection{$10^6 \msun$ Clusters}

In order to search for the parameters that may bring a cluster to the GC 
within several Myr via dynamical friction, we start our simulations with
a significantly more massive version of the Arches \& Quintuplet 
clusters.  Simulation~1 has a cluster having a total initial mass of
$M_{cl}=10^6 \msun$, a Plummer core radius of $r_c=0.86$~pc, and a 
circular orbit at $R=30$~pc.  Although we adopted a Plummer model for 
clusters for the sake of simplicity, equilibrium clusters located in an 
external tidal field are better represented by King models (King 1966).
The Plummer model does not have an intrinsic tidal limit, but the tidal radius, 
$r_t$, can be defined by
\begin{equation}
	r_t \equiv ( {M_{cl} \over 2 M_g} )^{1/3} R,
\end{equation}
where $M_g$ is the Galaxy mass enclosed inside $R$.  
The Plummer model with a ratio $r_t/r_c$ of 6 has core density and half-mass
radius comparable to those of the King model with $W_0=4$ (isotropic
King models have only one parameter, $W_0$, the concentration 
parameter; $W_0=4$ represents a moderate concentration).  The tidal radius at 
$R=30$~pc for a $M_{cl}=10^6 \msun$ cluster is 5.1~pc, so $r_t/r_c=6$
for Simulation~1, which represents an equilibrated cluster having
a moderate concentration and filling its tidal radius.

\begin{figure}[b]
\epsscale{1.15}
\plotone{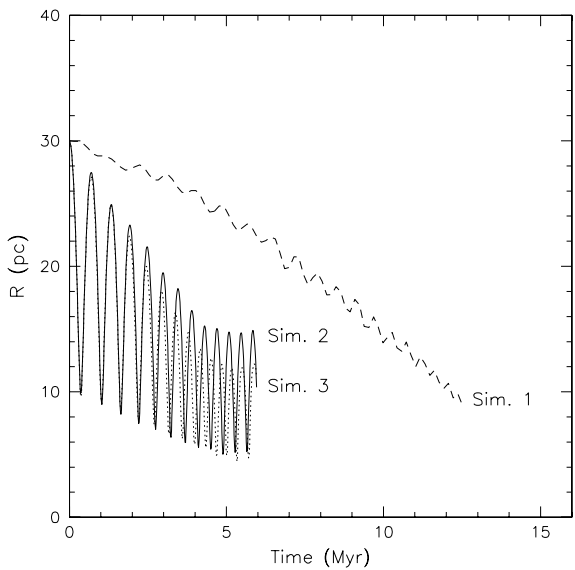}
\caption
{\label{fig:r30}Radial ($R$) evolution of the center of the cluster for
Simulations 1 ({\it dashed line}), 2 ({\it solid line}), and 3 ({\it dotted
line}).
}
\end{figure}

Figure~\ref{fig:r30} shows the evolution of Simulation~1.  It takes
$\sim 12$ Myr for the cluster to reach $R=9$~pc, where the cluster
completely disrupts, i.e., the core density ($\rho_c$) becomes smaller
than the local background density, $\sim 3 \times 10^3 \msun \,{\rm pc^{-3}}$.

The mechanism that triggers cluster formation, such as a cloud-cloud
collision, could dissipate some of the original energy of bulk motion into
a less ordered form such as turbulence or thermal energy.  This would lead
to an eccentric initial orbit for the cluster.  Since an eccentric orbit
would expose a cluster to a higher background stellar density at an earlier
phase, the cluster will experience relatively stronger dynamical friction
effects during a given time interval.  Here we test this notion by trying
in Simulation~2 a rather extreme example of an eccentric orbit: we set
the cluster to initially have a purely tangential velocity equal to half
of the circular velocity ($v_{circ}$) at the initial $R$.  All our eccentric
orbits in the present paper will have these conditions.  Simulation~2
otherwise has the same parameters as Simulation~1.  

This orbit of Simulation~2 plunges the cluster from $R=30$~pc to 10~pc in 
0.4~Myr.  Figure~\ref{fig:r30} shows the orbital motion of the cluster center. 
As a result of the eccentric initial orbit, the time to total disruption has
significantly decreased: we find that the density of the cluster core becomes
smaller than the background density at $\sim 5$~Myr. It is notable that,
in both Simulations 1 \& 2, the cluster core disrupts when it reaches
$R \simeq 10$~pc.

\begin{figure*}[t]
\epsscale{0.85}
\plotone{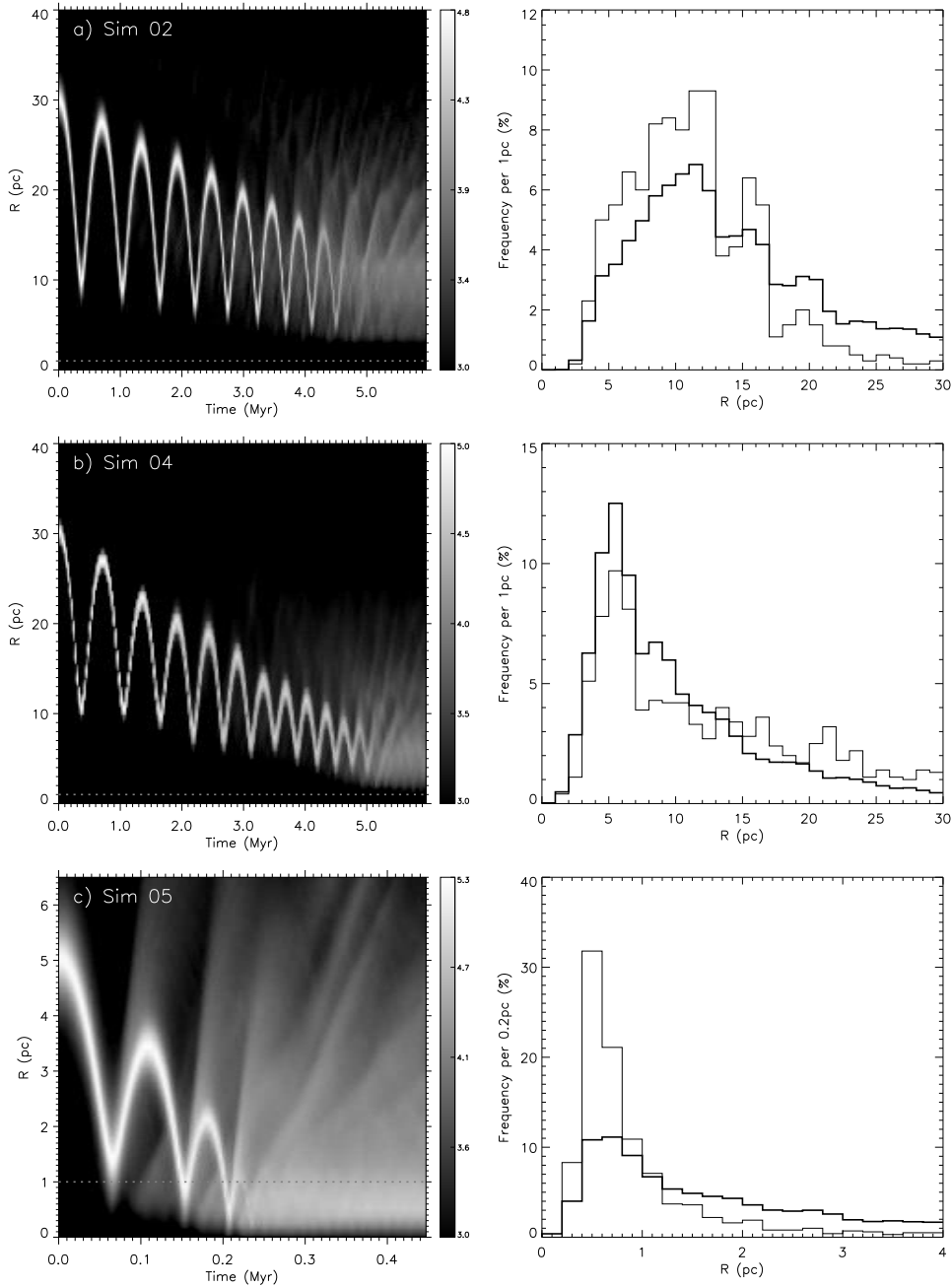}
\caption
{\label{fig:evolmap}Grey scale map for the temporal evolution of number
histogram for the radial distribution of cluster particles ({\it left panels})
and the same histogram for the final simulation step ({\it right panels}).
The grey scale bars next to the maps represent the scales of the density
in units of $M_\odot {\rm pc^{-3}}$.  In the right panels, thick lines
represent all stars in the cluster, and thin lines represent the central
1~\% stars at the beginning of the simulation.  The horizontal dotted line
in the map shows the location of $R = 1$~pc.
}
\end{figure*}

The center of the now unbound cluster continues to oscillate in $R$ even after
total disruption for a while ($t > 5$~Myr), but the minimum $R$ does
not vary significantly afterwards.  In fact, for this later period, our
definition of the density center of the cluster is simply the average
position of a group of particles that were bound to the cluster just
prior to disruption, and thus the center becomes meaningless after the
stars later become phase-mixed in their orbits.  At these later stages,
the full distribution of the cluster particles is more important for
understanding the fate of the cluster.  Figure~\ref{fig:evolmap}a displays
the evolution of the radial distribution of cluster particles (left panel),
and a histogram of the radial distribution at the final simulation step
(right panel).  After the disruption, the histogram peaks near $R=10$~pc
and has a rather distinct inner boundary at $R=3$~pc.  These results
show that the eccentric initial orbit can shorten $\tau_{fric}$ but it
does not help in allowing the cluster to penetrate deeper into the GC
before disruption.

The Chandrasekhar formula states that the drag force is inversely
proportional to the second power of the dispersion of the relative
velocity between the cluster and the background stars.  Thus, if the
background stellar system has a net rotation and the cluster rotates
in the same manner, then the relative velocity will be smaller and
the effects of dynamical friction will be larger.  Indeed, the central region
of the Galaxy does rotate (see \S~\ref{sec:galaxy}), although fairly weakly.
Simulation~3 is an attempt to assess the effect of such rotation on dynamical
friction.  Figure~\ref{fig:r30} shows that cooperative Galactic rotation only
slightly
enhances the friction effect, i.e., it brings the cluster inward more quickly
by just a fraction of a Myr.  We find that the final stellar distributions
of Simulations 2 and 3 are very much alike.

The three simulations discussed above disrupt at nearly the same $R$.
Aside from $M_{cl}$ and $R$, the core radius $r_c$ is the only common initial
parameter among the three simulations.  To investigate the dependence of the
final $R$ on the initial $r_c$, we perform Simulation~4, which initially has
$r_c=0.28$~pc, about 1/3 of that in Simulations 1--3.  This gives
$r_t/r_c$ = 19, and a central density of $3.8\times 10^6 \msun \,
{\rm pc^{-3}}$ (30 times that of the previous simulations).
This density is 1 to 2 orders of magnitude larger than the typical central density
of Galactic globular clusters.

The temporal progression of the density distribution in galactocentric radius 
for this simulation is shown in Figure~\ref{fig:evolmap}b.
Despite its relatively large initial core density, the cluster again disrupts
before reaching the central parsec.  However, the final stellar distribution
does peak at smaller $R$ ($\sim 5$~pc) than Simulations 1--3 ($\sim 10$~pc).
This is because clusters having a larger central density can survive in the
presence of larger background densities.  Nonetheless, it is clear that
the simulations discussed so far do not deliver massive stars into the
central parsec.

Now we move the initial $R$ closer to the GC.  Simulation~5 assumes the same 
cluster as Simulation~4, but placed initially in an eccentric orbit at $R=5$~pc.  
As shown in Figure~\ref{fig:evolmap}c, the cluster of Simulation~5 finally
reaches the central parsec and the core disrupts there.  However, after the
complete disruption, more than 60~\% of the cluster members are located
outside the central parsec at any given time.  On the other hand, the
histogram for the stars that initially constituted the central 1~\% of
the cluster shows that 70~\% of those stars end up projected
inside the central parsec at the final stage (thin line in the right panel).
This fraction increases to 90~\% when projected onto the sky.
Thus, if the progenitors of the \HeI stars are initially located at the
core of the cluster, simple binomial statistics give a probability
of $\sim 19$~\% for the 16 known \HeI stars to be found within the 
central parsec at any given moment after disruption.  This probability is 
large enough that we identify Simulation~5 as a candidate for the central 
parsec cluster.

The idea of an initial mass segregation of
heavier stars is consistent with some cluster formation models such as
the one by Murray \& Lin (1996), where encounters between cloudlets
increase the protostellar masses, and the one by Bonnell et al. (1997),
in which the deeper potential in the core causes stars there to accrete
relatively greater amounts of circumstellar material.
In addition, due to their compactness (large overall densities), clusters
formed near the Galactic center may undergo extremely rapid mass segregation
in the beginning of the evolution (Kim et al. 1999; this segregation is
not observed in our simulations because our cluster particles have the
same mass and more importantly, the internal dynamical evolution of the
cluster has been suppressed by giving non-negligible $\epsilon$ values
to the cluster particles.).  Thus, the initial mass segregation due to 
internal dynamics will preferentially bring massive stars to the central parsec.

To see if a cluster can reach the central parsec from a greater
distance, we initially locate the clusters of Simulations 6 and 7 
at $R=10$~pc.  Simulation~6 has the standard initial concentration,
$r_t/r_c=6$, while Simulation~7 has a higher concentration,
$r_t/r_c = 9.3$ (notice that Simulation~7 has the same $\rho_c$
as Simulation~5).  Figures \ref{fig:evolmap}d \& e show that
Simulation~6 results in a stellar disruption that leaves stars
widely distributed from 0.5 to 4~pc, and that the final phase of
Simulation~7 resembles that of Simulation~5.  90~\% of the central
1~\% of stars initially in the core of Simulation~7 are found within 
the central parsec
of the projected sky at the final stage.  As can be seen in the
comparison between Simulations 5 and 7, it appears that the final
distribution of stars is mainly determined by the initial $\rho_c$
(or, in more realistic clusters in which relaxation processes are 
possible, by the $\rho_c$ that is quickly established by mass segregation).
We identify Simulation~7 as another candidate for the central parsec
cluster.

There is a potential problem with $10^6 \msun$ cluster models.  While
their large mass causes the cluster to migrate inwards rapidly, it is
then very difficult to match the observed number of \HeI stars at the GC,
$\sim 16$ (Paumard et al. 2001).
Assuming that the progenitors of \HeI stars have an initial mass
$> 40 \msun$, the Salpeter initial mass function (IMF) with lower and
upper mass boundaries of 0.1 and $150 \msun$, respectively, will result
in $\sim 700$ \HeI stars out of a total mass of $10^6 \msun$.
To match the observed number, the power law slope of the IMF
needs to be as large as 3 (compared to 2.35 for the Salpeter IMF)
all the way down to $0.1 \msun$.  A slope of 3 represents an extremely
steep IMF, considering that even the disk IMF, which is thought to be
significantly steeper than the GC IMF (Morris 1993; Figer et al. 1999;
Kim et al. 2000), has a much flatter IMF for $<1 \msun$ (power law slope
of 2.1--2.3; Kroupa 2001 and references therein).  The assumed lower mass
cutoff at $40 \msun$ for \HeI stars is based on the suggestion by
Paumard et al. that the GC \HeI stars are in the luminous blue variable
phase or at the Wolf-Rayet stage.  If the lower mass cutoff for the helium
emission-line stars is, for example, $70 \msun$, the number of \HeI 
stars would be $\sim 250$.  On the other hand, if some of more massive 
\HeI stars, for example, those initially having
$>60 \msun$, have already undergone supernova explosions, the number
of remaining \HeI stars would be $\sim 350$ (the calculations by Schaller
et al. 1992 give 3.9 and 4.8~Myr for the lifetimes of solar-metallicity
[$z=0.02$] stars with initial masses of $40 \msun$ and $60 \msun$,
respectively).  However, these numbers are still significantly larger
than the observed number.\footnote{One could make this number even smaller
by assuming that the \HeI stars are not only a function of mass, but
also a function of the evolutionary phase, though.}

We have therefore investigated one more group of clusters:
clusters with a smaller initial $M_{cl}$ ($10^5 \msun$) starting at
closer $R$ ($\leq 10$~pc).

\begin{figure*}[t]
\figurenum{2 ({\it continued})}
\epsscale{0.85}
\plotone{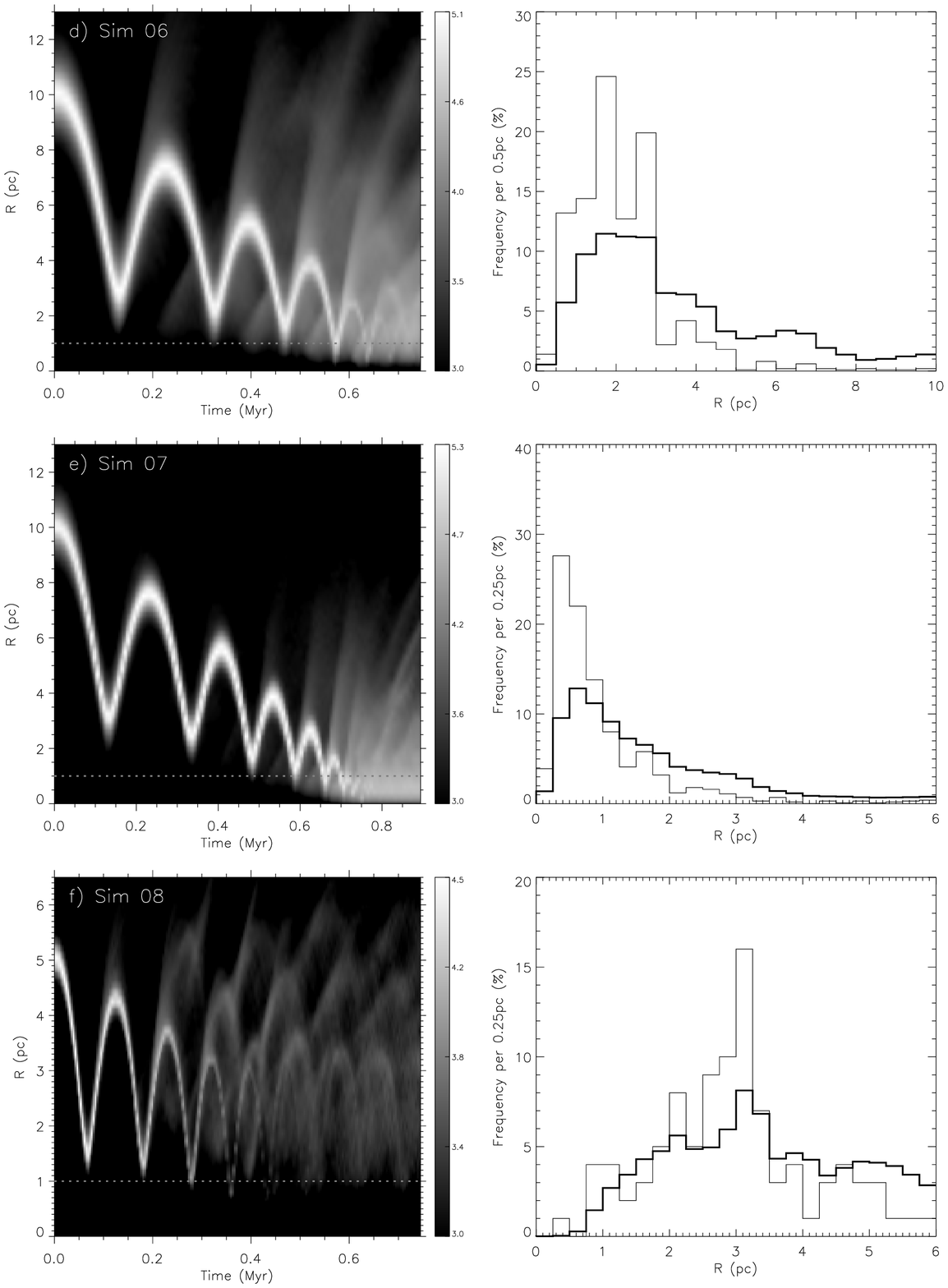}
\caption
{
}
\end{figure*}

\begin{figure*}[t]
\figurenum{2 ({\it continued})}
\epsscale{0.85}
\plotone{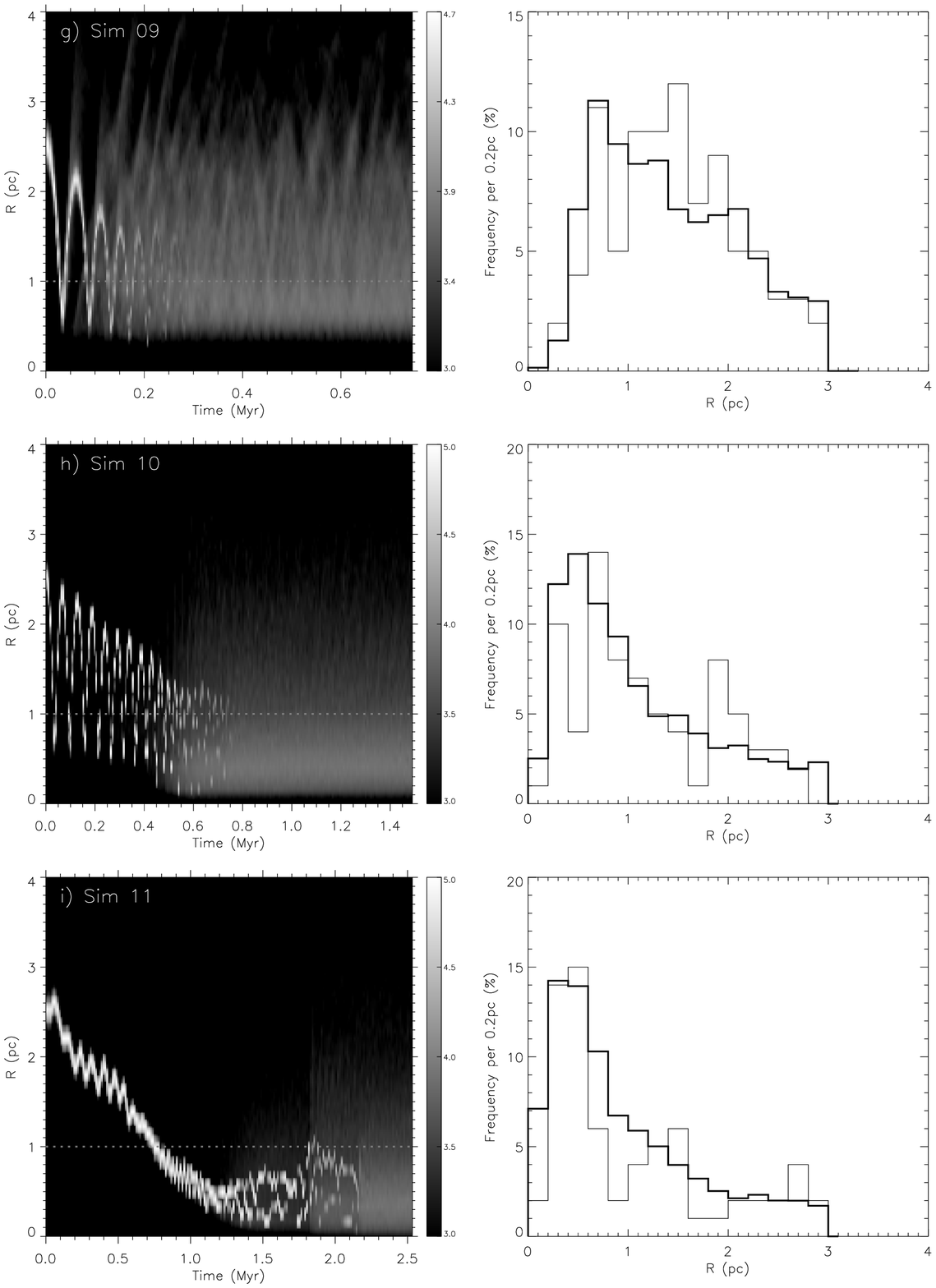}
\caption
{
}
\end{figure*}

\begin{figure*}[t]
\figurenum{2 ({\it continued})}
\epsscale{0.85}
\plotone{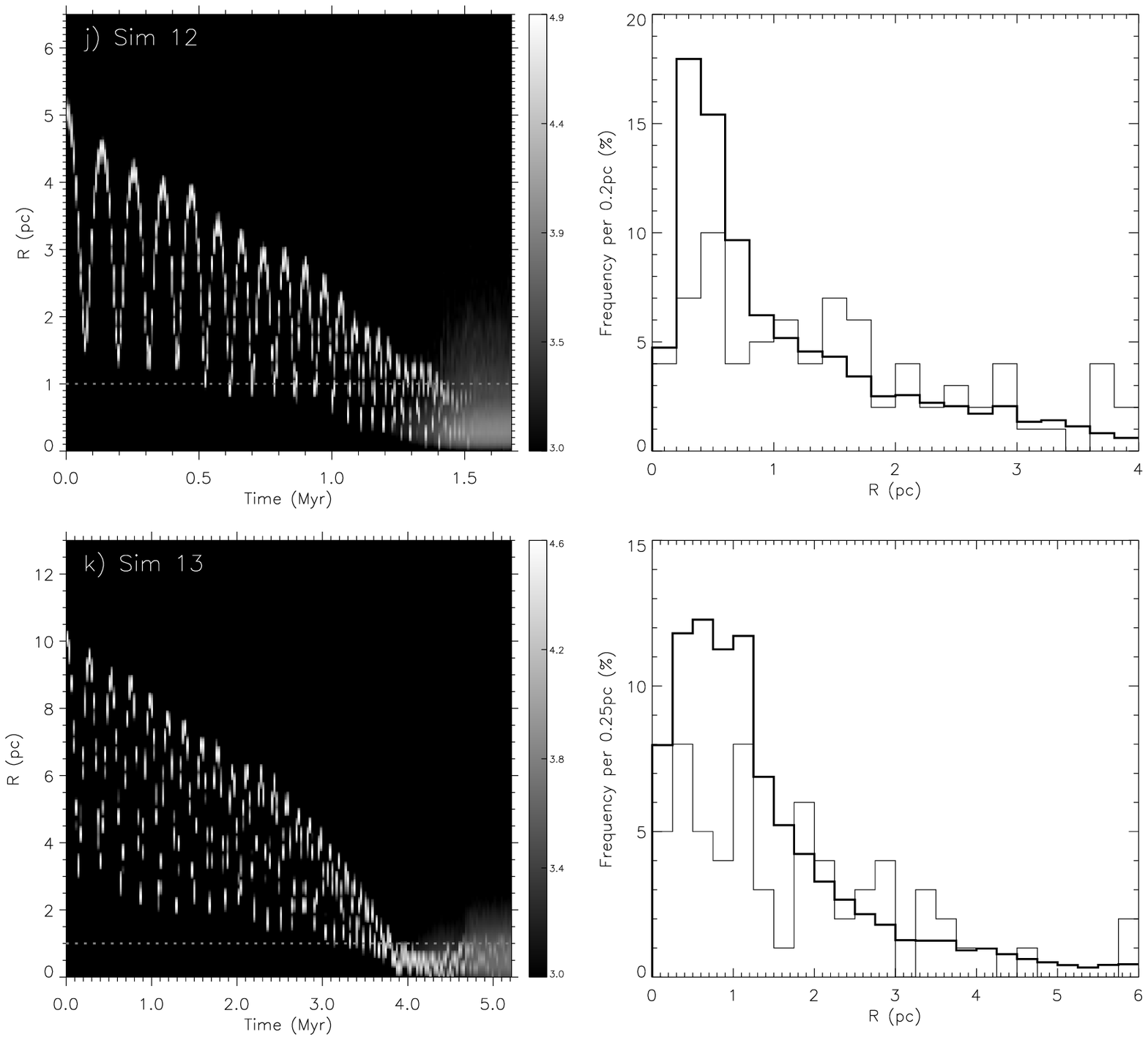}
\caption
{
}
\end{figure*}

\subsection{$10^5 \msun$ Clusters}

For a Salpeter IMF and mass limits of 0.1--$150 \,{ \rm M_\odot}$,
a cluster having $M_{cl}=10^5 \msun$ will yield $\sim 35$ stars with
masses between $40 \msun$ and $60 \msun$, which is now much closer to
the number of \HeI stars observed in the central parsec.

Simulations 8 and 9 have a moderate initial
concentration of $r_t/r_c=6$ (but relatively high $\rho_c$ of
$3.8 \times 10^6$ and $1.6 \times 10^7 \, {\rm M_\odot pc^{-3}}$),
and an eccentric initial orbit starting at $R=5$ and 2.5~pc,
respectively.  As shown in Figures \ref{fig:evolmap}f \& g, it takes
only a fraction of a Myr for these clusters to completely disrupt, and
the final stellar distributions are not confined to the central parsec.
While the peak of the final radial distribution of cluster stars in
Simulation~9 is just inside the central parsec, most particles of
Simulation~8 are located outside the central parsec.  These clusters,
although started at fairly small values of $R$, will not look like
the central parsec cluster, for which even higher initial $\rho_c$
will be required.

As an extreme case, we set Simulation~10 to have $r_t/r_c=12$ and an
eccentric initial orbit at $R=2.5$~pc.  This cluster will initially
have $\rho_c$ of $1.3 \times 10^8 \, {\rm M_\odot pc^{-3}}$, which
is an extremely high value.  Although it is highly questionable whether
a cluster could form with such a high central density, we note that
this assumption can be made in compensation for the fact that our
single-stellar-mass cluster models necessarily neglect the probably
important process of dynamical mass segregation for the Galactic
center clusters (Figer et al. 1999; Kim et al. 2000).
Mass segregation would concentrate toward the
cluster core exactly those stars which we would identify with the
HeI stars in the central parsec, so its neglect here leads to an
underestimate of the degree of concentration of massive stars toward
the center after disruption.  Our assumption of large initial cluster
density, is thus in part aimed at illuminating how mass segregation might
affect the hypothesis being investigated in this paper.
Figure~\ref{fig:evolmap}h shows that this cluster disrupts within only
1~Myr, inserting 49~\% of its members inside the central parsec.
When projected onto the sky, 64~\% of the stars are found in the
central parsec at the final stage.  This fraction is somewhat smaller
than those of Simulations 5 and 7, and the probability for 16 or so \HeI
stars to be found within the central parsec at any given epoch after 
disruption would be only $\sim 0.1$~\%.

Unlike Simulations 5 and 7, considering only the initial central 1~\%
does not increase the final concentration inside the central parsec for
Simulation~10.  Simulation~11 in Figure~\ref{fig:evolmap}i
shows that this same cluster can end up resembling the final stage of
Simulation~10 even with a circular initial orbit, although it takes
somewhat longer for it to undergo disruption.  65~\% of the stars
are found in the projected central parsec at the final stage of
Simulation~11.

Although the probability of finding about 16 \HeI stars inside the
central parsec, and none outside this radius, is small for Simulations 
10 and 11, the final
distributions from these simulations are quite concentrated at the GC.
In fact, our simulations, in which internal cluster dynamics are not
considered and all cluster particles have the same mass, may not
accurately represent the detailed structure of the cluster core, even
for cases in which we have augmented the initial core density to artificially
allow for mass segregation.
Thus, if the extremely compact cluster core consisting of massive
stars were to be better modeled, Simulations 10 \& 11 may then be able
to bring a much higher fraction of core stars down to the central
parsec.  We thus identify these two simulations as potential candidates
for the central parsec cluster.

Finally, we try this dense cluster model at greater initial $R$.
Simulations 12 and 13 have the
same initial central density as Simulation~10, but have greater
initial $R$'s, 5 and 10~pc, respectively.  Figures \ref{fig:evolmap}j \& k
show that the final distributions of Simulations 12 and 13 are similar
to that of Simulation~10, although the fraction of stars inside the
central parsec for Simulation~13 is slightly smaller ($\sim 43$~\%).
This confirms that it is the initial central density, not $R$, that
mainly determines the fate of the cluster, when only dynamical
friction is accounted for (i.e., when internal dynamics is not considered).
Note that it takes 1.5 and 4~Myr, respectively, for dynamical friction 
to bring the clusters in Simulations 12 and 13 into the central parsec.

Unlike our previous simulations, these time scales for dynamical friction
are comparable to the time scales by which clusters evaporate by relaxation
and stellar mass loss.  In particular, for an IMF slope of 2.35,
Kim et al. (1999) find that the cluster in Simulation 13 would
evaporate within $\sim 4$~Myr, if it stayed at its initial $R$, 10~pc.
It is difficult to estimate how important the relaxation will be
when the dynamical friction makes the cluster rapidly lose its mass
by moving it into a stronger tidal potential (so that the tidal radius
becomes smaller).  However, it is certain that the
cluster models considered in the present study are more enduring than
more realistic models might be, and consequently serve as a useful
limiting case.

\section{DISCUSSION}
\label{sec:discussion}

We have performed numerical simulations for dynamical friction of
star clusters at the Galactic center with $M_{cl} = 10^5$--$10^6
\msun$ and $R = 2.5$--$30$~pc.

We find that extremely massive versions of the Arches and Quintuplet
clusters (Simulations 1--4; $M_{cl}=10^6 \msun$ \& $R = 30$~pc) disrupt
before reaching the central parsec.
When initially located at $R \leq 10$~pc,
clusters having $M_{cl}=10^6 \msun$ can reach the central parsec within the
estimated age of \HeI stars, $\sim 5$~Myr, (Simulations 5 \& 7), but the
final stellar distribution is spread beyond the central parsec.
However, stars initially located at the core of the cluster
(as is likely to be the case for \HeI stars, which
are massive enough to sink rapidly to the cluster core soon after
formation) tend to be more concentrated in the central parsec at the
final stage: of the initially innermost 1\% of the cluster stars, 90~\% 
of them end up projected onto the central parsec in Simulations 5 \& 7, which
we identify as candidates for the central parsec cluster.  

Smaller clusters having a higher central concentration and initially
located relatively near the GC
(Simulations 10--13; $M_{cl}=10^5$, $R=2.5-10$~pc, \& $\rho_c = 1.3
\times 10^8 \msun pc^{-3}$) are found to be {\it potentially} another
set of candidates for the central parsec cluster.  Unlike the $10^6 \msun$
clusters, however, the stars initially located in the innermost 1\% of the
cluster core of these ``low-mass'' clusters have a 
final concentration to the central parsec of the GC similar to that of 
the entire population of cluster stars, $\sim 50$~\%.  Also, one must
keep in mind that even these ``low-mass'' clusters are an order of magnitude
more massive than the massive Arches and Quintuplet clusters.

We have found that the $10^5$ and $10^6 \msun$ clusters have too many 
\HeI stars compared to the number found in the central parsec and the
surrounding volume.
A $10^4 \msun$ cluster composed of only stars, on the other hand, would
have the appropriate number of \HeI stars, but it would not experience
significant dynamical friction before complete cluster evaporation, as the
frictional deceleration is expected to
scale as $M_{cl}^{-1}$.  Thus it appears that, in order for dynamical
friction to be responsible for the central parsec cluster, the total
cluster mass should be $\ga 10^5 \msun$, and some phenomenon must be
invoked in order to reduce the number of \HeI stars. Perhaps most of 
them have disappeared as supernovae, while others have been lost by
evaporation at a rate which exceeds that in our models.  The evaporation
hypothesis, however, leaves us with the embarrassment that large numbers
of massive \HeI stars should have been detected (Figer 1995).  Another
possible explanation is that the lower mass limit for \HeI stars is higher
than is usually considered to be the case.  Finally, the possibility has
been raised that a significant number of massive stars have merged in the
dense stellar core to form an intermediate-mass black hole (Portegies Zwart 
\& McMillan 2002; Hansen \& Milosavljevic 2003).  

The large cluster mass required for the hypothesis that dynamical friction
is responsible for bringing the central young cluster into the central
parsec is not a problem in principle.  While there exists no known, young
cluster near the center of our Galaxy having a mass within an order of 
magnitude of that required, such massive clusters have been found in many
other -- usually starburst - galaxies, and are usually referred to as 
super star clusters.  

Finally, we note that, if the dynamical friction hypothesis is applicable
to luminous young stars in the central parsec, then the collapsed core of
the original cluster might still be present there.  The finding of a very
compact configuration of massive stars in the central parsec would strengthen
this hypothesis considerably.  

In conclusion, we have identified a few simulations that can be regarded
as candidates for the origin of the central parsec cluster.  However,
the required conditions are rather extreme.  While it is more probable
for a massive cluster ($10^6 \msun$) to reach the central parsec before
disruption, one needs a finely-tuned set of parameters to observe only
$\sim 16$~\HeI stars out of the whole original mass and to have them be
concentrated almost entirely into the central parsec.  Less massive clusters
($10^5 \msun$) might have less problem with the \HeI star count, but
require either an extremely high initial central density (inherently or
via relaxation) or a rapid segregation of massive stars in the cluster
core.  Consideration of internal dynamics and mass spectrum in the cluster
will enable core collapse and mass segregation and thus tend to raise the
probability for the cluster core to reach the central parsec intact, but these
same effects also make the cluster evaporate more quickly and thus 
weaken the drag from dynamical friction, thus lowering that probability.
The internal dynamics of the cluster are more important for
a more compact cluster located at a larger $R$.  We have artificially
and partially explored one of the consequences of the internal cluster
dynamics by trying some models with extremely high central densities.
However, numerical simulations that can handle both the galactic scale
(dynamical friction) and the cluster scale (internal cluster dynamics)
are still needed to show the exact role of the internal dynamics for
the structural evolution of our cluster models for the time intervals
considered here.

\acknowledgements
S.S.K. thanks Michael Fellhauer, Hyung Mok Lee, David Merritt, Andres Meza,
and Martin Weinberg for helpful discussion.  M.M. acknowledges interesting
and fruitful conversations with Oertwin Gerhard and Simon Portegies-Zwart.
We are grateful to Gaber Mohamed at Academic Technology Services (ATS) of
UCLA for his kind assistance with issues on supercomputing.  We have used
computer facilities at ATS of UCLA, San Diego Supercomputer Center, and
Maui High Performance Computing Center of the University of New Mexico.
Work by S.S.K. was supported by the Astrophysical Research Center for the
Structure and Evolution of the Cosmos (ARCSEC) of Korea Science and Engineering
Foundation through the Science Research Center (SRC) program.


\end{document}